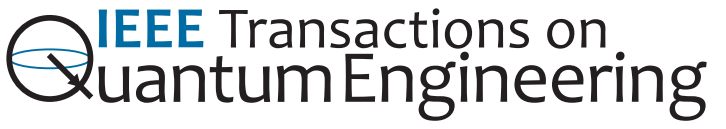

# Network-Assisted Collective Operations for Efficient Distributed Quantum Computing

**IAGO FERNÁNDEZ LLOVO, GUILLERMO DÍAZ-CAMACHO, NATALIA COSTAS LAGO, AND ANDRÉS GÓMEZ TATO**

Galicia Supercomputing Center (CESGA), 15705 Santiago de Compostela, Spain

Corresponding author: Iago Fernández Llovo (e-mail: ifllovo@cesga.es).

This work was supported in part by the Ministerio de Ciencia, Innovación y Universidades (MICINN) through the European Union NextGenerationEU recovery plan, the Galician Regional Government through “Planes Complementarios de I+D+I con las Comunidades Autónomas” in Quantum Communication under Grant PRTR-C17.I1, in part by the Galicia Supercomputing Center (CESGA) for providing access to the FinisTerrae III supercomputer with financing from the Programa Operativo Plurirregional de España 2014-2020 of the European Regional Development Fund (ERDF) under Grant ICTS-2019-02-CESGA-3, and in part by the European Union through Galicia Supercomputing Center (CESGA) to its Qmio quantum computing infrastructure, through the Programa Operativo Galicia 2014-2020 of ERDF–REACT EU, as part of the European Union’s response to the COVID-19 pandemic. Iago Fernández Llovo acknowledges support from Xunta de Galicia through Grant IN606B-2025/027, as significant parts of the peer review process were done after this funding was awarded. *(Iago Fernández Llovo and Guillermo Díaz-Camacho contributed equally to this work.)*

**ABSTRACT** Distributed quantum computing relies on coordinated operations between remote quantum processing units (QPUs), yet most existing work either assumes full connectivity, unrealistic for large networks, or relies on entanglement swapping. To mitigate the overhead of communication, we propose a scheme for the distribution of collective quantum operations among remote QPUs by exploiting distributed fan-out operations to a central node in network architectures similar to those used for high-performance computing, which requires only preshared entanglement, local operations, and classical communication. We show that a general diagonal gate can be distributed among any number of nodes and provide the ebit cost bounds. For a single distributed multicontrolled gate, this amounts to a single additional Bell pair over the theoretically optimal calculation with all-to-all preshared entanglement, demonstrating better scalability when compared to current proposals based on entanglement swapping through a network. We provide a recipe for the lumped distribution of gates, such as arbitrarily sized Toffoli and multicontrolled Z and $R_{zz}(\theta)$ gates. Finally, we provide an exact implementation of a distributed Grover’s search algorithm using this protocol to partition the circuit, with Bell pair cost growing linearly with the number of Grover iterations and the number of partitions, and show how these techniques can be applied to other algorithms, such as quantum approximate optimization algorithm. Our results show that alternative approaches to entanglement swapping can provide major benefits in distributed quantum computing, pointing to promising avenues for future research.

**INDEX TERMS** Distibuted quantum computing (DQC), Grover’s search algorithm, quantum approximate optimization algorithm (QAOA), quantum networking, teleportation.

## I. INTRODUCTION

Achieving quantum systems with thousands or millions of qubits [1], [2], [3], essential for fault-tolerant algorithms, such as Shor’s factoring and Grover’s search [4], [5] or quantum simulation [6], remains a key challenge in quantum computing. Distributed quantum computing (DQC) addresses the scalability issue by integrating multiple smaller quantum processors into a single, unified system with higher computational power [7], [8]. The distributed approach has demonstrated advantages in scalability, performance, robustness, and cost in classical high-performance computing (HPC) [9], benefits that DQC systems can also leverage by sharing resources between nodes [7], [8].

Entanglement is the cornerstone of quantum computing, so distributed systems require devices capable of producing entanglement between them [8], [10]. Typically, this can be achieved via the generation of Bell states, demonstrated experimentally in multiple quantum platforms, including diamond nitrogen vacancy (NV)-centers [11], [12], [13], [14], [15], superconducting circuits [16], [17], and trapped ions [18], [19], [20], with fidelities reaching 88% over 230 m [21]. Once entangled states are available, quantum teleportation can facilitate the propagation of quantum information. Two main types of quantum teleportation are known: state teleportation (teledata), transferring quantum states between devices [12], [22], [23], [24], and gate teleportation (telegate), enabling the remote execution of quantum gates [25], [26], [27], [28]. Particular algorithms may be better suited to specific network topologies, especially when considering realistic constraints, such as Bell pair generation time and communication latency. The concepts of teledata and telegate were first introduced by Van Meter et al. [29], where various arithmetic operations across different network topologies were analyzed, taking into account the costs associated with entanglement generation and communication.

In the case of bipartite entanglement, at least a preshared Bell pair, or ebit (from entangled qubits), must be consumed to perform teleportation operations, or a Greenberger–Horne–Zeilinger (GHZ) state for multipartite entanglement [30]. Therefore, a number of ebits are required to complete a distributed algorithm. Note that this amount may be larger than the physical number of communication qubits—qubits that can directly communicate in a network to generate entanglement via transduction or other means [8]—as a single communication qubit may be reused multiple times throughout a computation by resetting its state and regenerating entanglement [31]. Moreover, communication qubits can be implemented using different matter qubits than data qubits, e.g., two different species in trapped-ion systems [32], [33] or the electron and nuclear spins in NV-center systems [15]. Recently, the telegate method has been used for the experimental demonstration of two-qubit Grover's algorithm across two nodes with an optical link in dual-species ion traps, using $^{43}Ca^{+}$ for data qubits and $Sr^{+}$ for communication qubits [33].

To generate and use entanglement across a network of quantum devices, much work has been focused on the entanglement swapping model [8], [10]. By generating Bell pairs between intermediary nodes and performing Bell state measurements (BSMs), end nodes become entangled without direct information exchange [8], [10], [34], [35], [36], [37]. This is illustrated in Fig. 1, where the gate-based entanglement swapping scheme for three nodes is shown. Entanglement swapping was initially demonstrated with polarization-entangled photons [38] and has since been achieved in other quantum systems, e.g., separated solid-state quantum memories [39] or NV-centers in a multinode teleportation network [15]. While this model enables the generation of Bell pairs between any two devices linked by quantum routers, switches, or repeaters [8], [15], [37], [40], DQC networks must be optimized for latency, throughput, and robustness, while reducing the communication burden [8]. Therefore, while entanglement swapping may be adequate for a completely unstructured and unknown network, such as a future quantum Internet [40], alternative strategies may be better tailored to DQC.

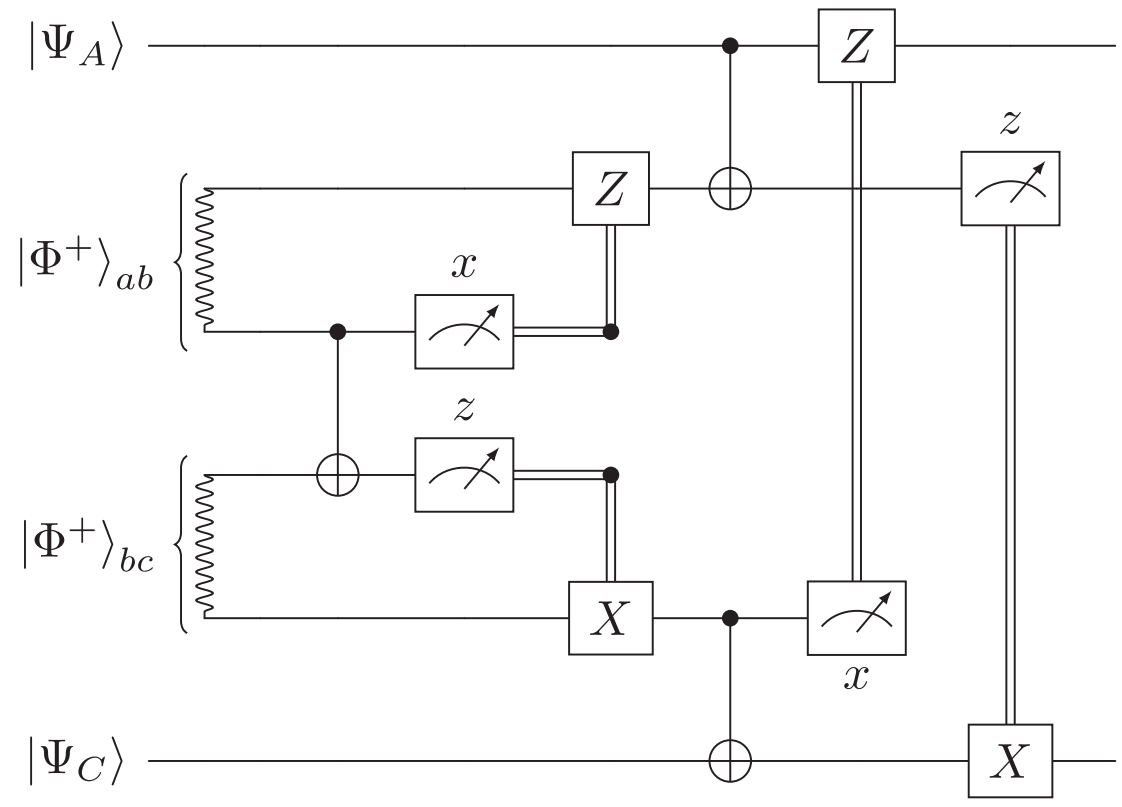

**FIGURE 1.** Entanglement swapping scheme in the gate-based quantum computing model. Alice and Charlie each share Bell pairs $|\Phi^+\rangle$ with Bob (squiggly lines). A BSM at Bob's node projects Alice and Charlie into an entangled Bell pair, without direct quantum or classical communication between them. Two bits of classical information (double lines) are required to conditionally correct the resulting state. The Bell pair can then be measured—hence consumed—in a quantum teleportation protocol (in the picture, a telegate CNOT).

In this article, we introduce a mechanism for teleporting large diagonal quantum gates using network devices as accelerators for collective operations involving many nodes in a network. This scheme relies on nonlocal fan-out operations, allowing the lumped execution of multiple diagonal gates in a single round. We show analytically that the ebit consumption is only one above the theoretical minimum and apply this scheme to distributed Grover's search. Furthermore, we also discuss its relevance for other algorithms, such as quantum approximate optimization algorithm (QAOA), demonstrating that a clusterized model can be easily distributed in a star network following this method.

This approach demonstrates that entanglement swapping may be suboptimal in some applications and that ebit costs can be reduced by leveraging network devices beyond simple entanglement distribution. We show that relevant gates, such as arbitrarily sized multicontrolled-Z (MCZ) gate, multicontrolled phase gates $MCR_z(\theta)$, or collective rotation gates $R_{zz}(\theta)$, can be teleported efficiently following this model. Our approach is well suited to star topologies, with a single central network node (i.e., router), with which all other nodes have direct connectivity, as a first step toward building a toolset for a DQC model where network devices participate in the computation instead of simply distributing entanglement.

## II. PRELIMINARIES

### A. NETWORKING IN HPC

Quantum networks for DQC must implement all the experimental apparatus required to generate entanglement between arbitrary nodes. DQC network constraints resemble those of HPC systems, where network architectures must achieve low and predictable latency and nonblocking all-to-all connectivity [9]. In contrast, unstructured networks proposed for quantum Internet applications [34], [35], [36] are unsuitable for current supercomputers.

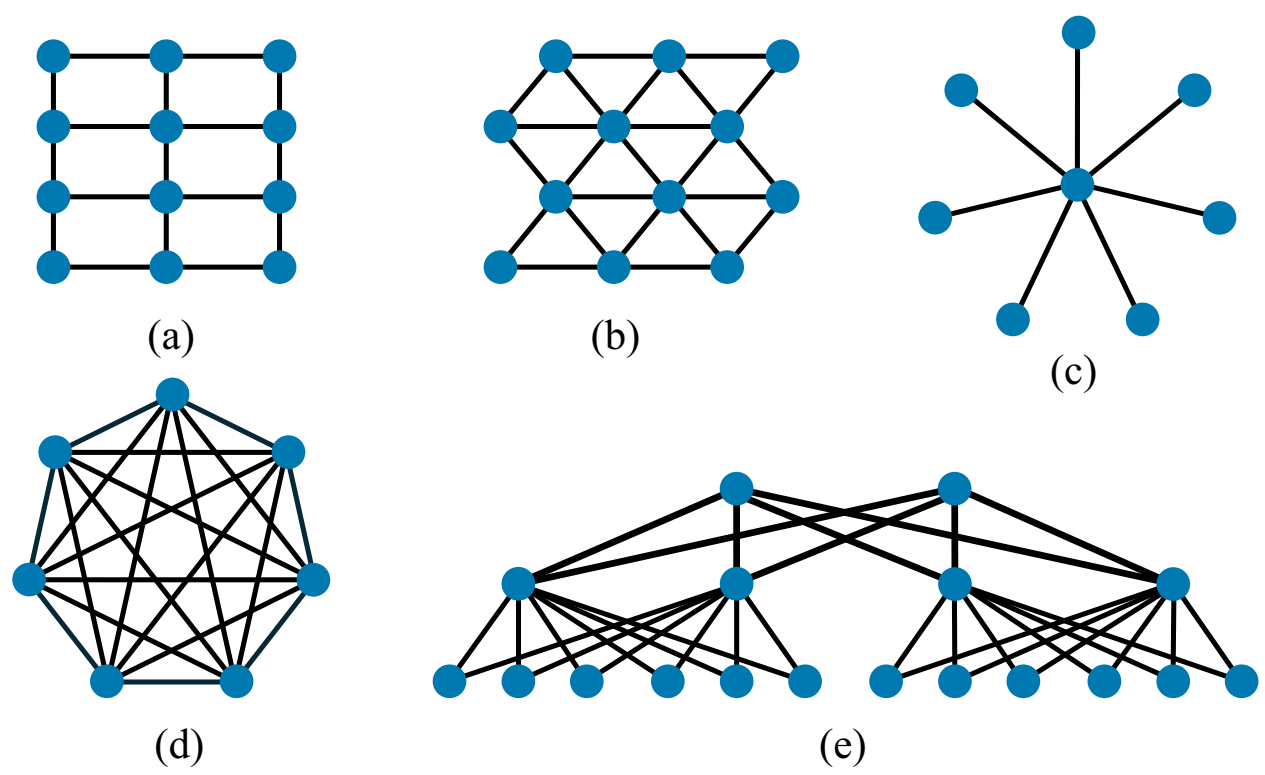

**FIGURE 2. Examples of network topologies. (a) and (b) Four and six first neighbor connectivity with $n = 2N$ and $3N$ connections per node, (c) star topology, with $n = N$ connections, (d) all-to-all connectivity, with $n = \binom{N}{2}$ connections, and (e) three-tier network, where thicker lines illustrate higher bandwidth links.**

Structured network topologies include first-neighbor configurations, such as square or hexagonal lattices [see Fig. 2(a) and (b)] and star topologies, where all communications across network nodes pass through a central node [see Fig. 2(c)]. In first-neighbor network topologies, path lengths—defined as the number of hops required to reach a destination node—vary significantly. So, communication from opposite sides of the network requires numerous hops, leading to high and variable latency. Star networks ensure a path length of two but introduce a single point of failure and a potential bottleneck under heavy load. While all-to-all networks [see Fig. 2(d)] meet HPC requirements, they scale poorly, requiring $\binom{N}{2}$ interconnections for $N$ nodes.

Due to the limitations of simple topologies, more complex, multitier architectures, such as Clos, fat-tree, and spine-leaf networks, have been widely adopted in HPC data centers, as they alleviate the burden of $O(N^2)$ connection growth [41], [42] while maintaining high-throughput, and form the basis of switched fabrics, such as InfiniBand [43]. Fig. 2(e) illustrates a three-tier network. In such architectures, any node can reach another in a few hops, ensuring low and predictable latency even under high load, while providing nonblocking, robust operation with no single points of failure. These networks have been shown to provide optimal connectivity [41], [42], so they could similarly benefit scalable quantum systems. Moreover, multitier networks are inherently dynamic, as nodes can be added or removed without disrupting the rest of the system, enabling gradual scalability.

### B. COLLECTIVE OPERATIONS FOR DQC

In classical computing, collective operations refer to parallel programming routines involving multiple nodes. So, implementing quantum gates as collective operations, i.e., quantum gates involving qubits physically located in multiple nodes across the network, can help scaling distributed quantum computers without a loss of universality. While single- and two-qubit operations are widely discussed, many quantum algorithms rely on multiqubit operations, e.g., the diffuser (and often, the oracle) in Grover's search algorithm [5]. Notably, Toffoli (and thus CCZ) plus Hadamard gates form a universal gate set [44]. Although any unitary gate can theoretically be implemented using teleportation [25], [45], the optimal implementation and ebit cost remain unclear.

Let us first discuss how collective operations can be implemented across $N$ nodes. One approach is teleporting the quantum state of all involved qubits to a single node, where any unitary $U$ can be applied locally before teleporting the states back to their original locations, known as teledata. A big disadvantage of this approach is that a large enough quantum processing unit (QPU) must be available, with at least twice the number of qubits of the largest operation possible for Bell pair creation and distribution. The state must be teleported back to the original location to free the computing resources, or else schedule the teleportation of the individual qubits after other local operations have been performed.

Another approach, known as telegate operations, can also be used to implement distributed operations, such as CNOT and Toffoli gates [46], [47] and teleportation of the control in controlled unitaries [48], [49]. Assuming the minimum number of hops through a network to be 2 (corresponding to a star network topology), and given that entanglement can be distributed beforehand, the tally comes out as $4N$ ebits required for teledata operation and $2N$ for telegate operation. More critically, in both cases, one of the QPUs must allocate additional ebits for the calculation, increasing both the ebit-bandwidth required in this node and potentially reducing the number of effective data qubits available for the rest of the algorithm.

In Section III, we will show how $k$-node collective operations can be implemented with $k$ ebits and which type of gates can be implemented following this scheme. We also demonstrate how Grover's algorithm for unstructured search can benefit from distributing large collective operations.

### C. FAN-OUT AND CAT-ENTANGLER

Let us now discuss what tools are at our disposal to perform parallel operations on a quantum state $|\psi\rangle = a_0 |0\rangle + a_1 |1\rangle$. We would like to perform a hypothetical quantum broadcast operation, such as

$$|\psi\rangle |0\rangle^{\otimes n-1} \rightarrow |\psi\rangle^{\otimes n} \tag{1}$$

to then recombine the outputs

$$U_1 \otimes U_2 \otimes \cdots \otimes U_n |\psi\rangle^{\otimes n} \rightarrow U_1 U_2 \ldots U_n |\psi\rangle |0\rangle^{\otimes n-1}. \tag{2}$$

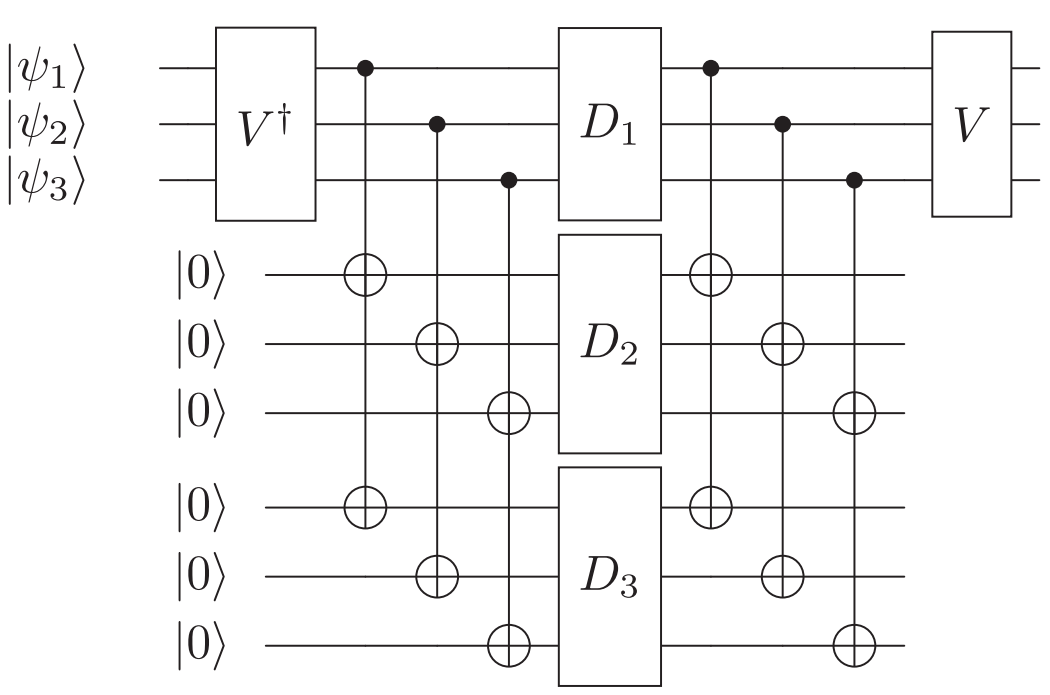

**FIGURE 3.** **Parallelization of the multiqubit $U_i$ gates, which are diagonalized as $U_i = V^\dagger D_i V$. A fan-out operation can be used to expand the qubits and apply the diagonal gates $D_i$ in parallel, instead of consecutively applying $U_1U_2U_3$ over the original qubits.**

This procedure, which would enable the application of $U_{1\cdots n}$ simultaneous unitary gates $U_1 \otimes U_2 \otimes \cdots \otimes U_n$, is, however, strictly forbidden for unknown states by the quantum no-cloning theorem [50]. Nevertheless, entanglement allows us to perform a similar operation in the quantum realm: a state can be expanded across multiple qubits through a fan-out operation [51]

$$(a_0 |0\rangle + a_1 |1\rangle) |0\rangle^{\otimes n-1} \rightarrow a_0 |0\rangle^{\otimes n} + a_1 |1\rangle^{\otimes n}.$$

The resulting state has "opened" as a handheld fan into several qubits—known as a generalized GHZ or *cat-state* [30]. Quantum operations can then be applied in parallel on this state. The resulting qubits can then be recombined or "closed" with the reverse operation of the fan-out (or simply, *fan-in*), returning the desired $U_1U_2 \ldots U_n |\psi\rangle$. Parallelization has a single requirement: that gates commute pairwise [51]. The basis in which the gates are all diagonal must be known, and the transformation must be performed outside of the fan-out operation.

A tradeoff must be made between circuit depth and number of ancillae required when parallelizing operations following this scheme: for single-qubit gates, only one ancilla per gate is required, but for $N$-qubit gates, each participating qubit requires to be expanded individually to $N$ respective ancillae, as shown in Fig. 3.

For instance, the action of three $U_i = Rz(\theta_i)$ rotation gates can be parallelized

$$\begin{aligned}
&(a_0 |0\rangle + a_1 |1\rangle) |00\rangle \xrightarrow{\text{fan-out}} \\
&a_0 |000\rangle + a_1 |111\rangle \xrightarrow{U_1 \otimes U_2 \otimes U_3} \\
&a_0 |000\rangle + a_1 e^{i\theta_1} e^{i\theta_2} e^{i\theta_3} |111\rangle \xrightarrow{\text{fan-in}} \\
&(a_0 |0\rangle + a_1 e^{i\theta_1} e^{i\theta_2} e^{i\theta_3} |1\rangle) |00\rangle
\end{aligned} \tag{3}$$

accumulating the phases after the fan-out/fan-in process. To parallelize multiqubit gates, each qubit has to be expanded to its respective ancillae, as shown in Fig. 3. In general, finding the common basis for a series of multiqubit gates is difficult, but this issue is alleviated when parallelizing controlled gates. For any controlled unitary $CU_i$, the control part of the gate is always diagonal in the computational basis, irrespective of the shape of $U_i$. The effect on the control qubit is a phase-kickback $e^{i\theta_i}$, and the gates can be parallelized without transformation.

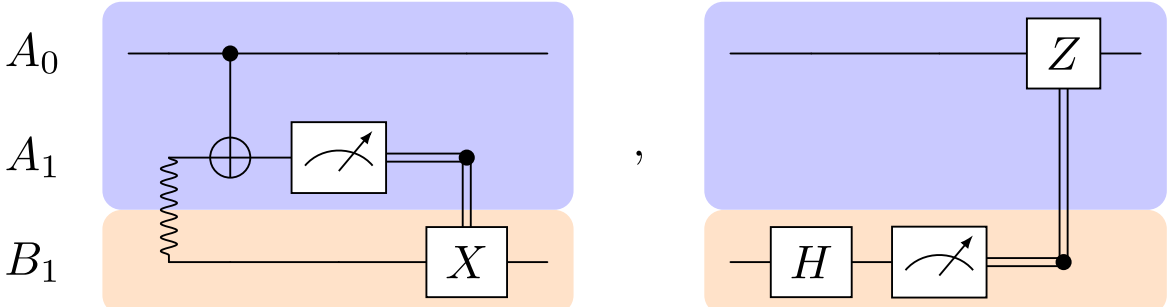

**FIGURE 4.** **(Left) Cat-entangler and (right) cat-disentangler protocols as described in [30], where qubits $A_{0,1}$ are physically in the same node, and $B_1$ is a remote qubit pre-entangled with $A_1$, forming a Bell pair (squiggly line). These are equivalent to fan-out and fan-in between nodes.**

In Fig. 3, the $N$ qubit fan-out gate is presented as a control gate with $N-1$ targets. An unbounded (i.e., not limited in the number of qubits it acts) fan-out gate can be constructed with a simple ladder of CNOT gates controlled by the original qubit and ancilla qubits initialized in $|0\rangle$ as targets [51]. However, there are more efficient implementations than the straightforward cascade of CNOT gates: as the resulting state is a generalized GHZ state, efficient implementations of GHZ states could also be used, resulting in lower depths, e.g., logarithmic depth succession of CNOT gates [52] or constant depth of six using mid-circuit measurements and feedforward control [53]. Furthermore, fan-out gates can be more easily executed in some hardware by means of more sophisticated, native interactions, e.g., global interactions, such as the global Molmer–Sörensen (GMS) gate in ion traps [54], [55] or power–law interactions in Rydberg atoms or diamond NV-centers [56]. Interestingly, a remote execution of the three-qubit fan-out gate without decomposition into two-qubit gates was experimentally demonstrated in [57] to create these generalized GHZ states between several nodes.

The fan-out/fan-in protocol can be extended for remote multiqubit operations in DQC, sharing the generalized GHZ state between separate devices. This protocol, referred to as cat-entangler/cat-disentangler by Yimsiriwattana and Lomonaco [30], was used for a distributed version of Shor's algorithm [58], further improved by Van Meter et al. [29] when taking into account network topology and focusing more on teledata. Häner et al. [59] also introduced a similar concept, rebranded as fanout/unfanout in their quantum message-passing interfaces (MPI) proposal. A further extension of this protocol is used in [60] and [61] as starting process and ending process corresponding to the cat-entangler and the cat-disentangler, respectively. For simplicity, the term fan-out will be used interchangeably with cat-entangler in this text when referring to operations between nodes. These two operations are sketched in Fig. 4.

Unlike protocols designed to parallelize operations in a monolithic QPU, the goal here is to facilitate the application of gates without requiring a direct connection between devices. This is similar to executing remote gates between qubits on a QPU with limited connectivity, assuming that one has qubits in the $|0\rangle$ state to act as ancillae. The protocol involves expanding the state to ancillary qubits shared among the parties, locally applying the desired gate, and then applying fan-in to disentangle the qubits. This process uses Bell pairs as the resource for teleporting the fan-out and fan-in gates. While it may seem necessary to teleport two separate CNOT gates, the entire process can be performed with the cost of only one ebit and two classical bits [46]. In the first step, one ebit and one classical bit are shared between the remote parties. In the second step, only an additional classical bit is required to complete the operation.

Given a quantum state at one node, $|\psi\rangle = a_0|0\rangle + a_1|1\rangle$, and a preshared Bell pair between the two nodes, $|\Phi^+\rangle = \frac{1}{\sqrt{2}}(|00\rangle + |11\rangle)$, it is possible to combine both into a larger cat-state, $\frac{1}{\sqrt{2}}(a_0|00\rangle + a_1|11\rangle)$. This effectively distributes the amplitudes $a_0$ and $a_1$ of the original state across the two nodes, enabling controlled operations to be performed at the second node. As described in [30], the process begins with the product state

$$\frac{1}{2}\left[a_0|0\rangle(|00\rangle + |11\rangle) + a_1|1\rangle(|00\rangle + |11\rangle)\right] \tag{4}$$

where the first qubit corresponds to $|\psi\rangle$ and the other two qubits correspond to the Bell pair. A CNOT gate is then applied between the first qubit (control) and one of the Bell pair qubits (target), resulting in

$$\frac{1}{2}\left[a_0(|000\rangle + |011\rangle) + a_1(|110\rangle + |101\rangle)\right] \tag{5}$$

where the CNOT gate has been applied on the second qubit. This qubit is then measured, resulting in two possible outcomes. If the measurement result is 0 (underlined below), then the state collapses to

$$\frac{1}{\sqrt{2}}\left(a_0|0\rangle|\underline{0}0\rangle + a_1|1\rangle|\underline{0}1\rangle\right) \tag{6}$$

or, if the measurement result is 1, then to

$$\frac{1}{\sqrt{2}}\left(a_0|0\rangle|\underline{1}1\rangle + a_1|1\rangle|\underline{1}0\rangle\right). \tag{7}$$

After this measurement, either the state in (6) or (7) is obtained, representing a mixed state. A pure state can be recovered by using the measurement outcome to classically control an $X$ gate in the third qubit, ensuring the outcome is the desired cat-state $\frac{1}{\sqrt{2}}(a_0|00\rangle + a_1|11\rangle)$. This can be generalized with larger GHZ states of $m$ qubits $|\mathrm{GHZ}_m\rangle = \frac{1}{\sqrt{2}}(|00\ldots0\rangle + |11\ldots1\rangle)$. In that case, the measurement in the second qubit controls $X$ gates in all the remaining qubits to correct them.

The second process, cat-disentangler, allows for the fan-in operation to be applied deterministically on the remote qubits, requiring only local operations and classical communications, after some diagonal operation has been performed remotely using the shared state. Starting with the cat-state $\frac{1}{\sqrt{2}}(a_0|00\rangle + a_1|11\rangle)$, a Hadamard gate is applied on the remote qubit (or qubits, in the case of a larger GHZ state). This operation transforms the state into

$$\frac{1}{2}\left[a_0(|00\rangle + |01\rangle) + a_1(|10\rangle - |11\rangle)\right]. \tag{8}$$

Measuring the remote qubit results in two possible outcomes for the remaining qubit. If the result of the measurement is 0, then the state becomes

$$\frac{1}{2}\left(a_0|0\rangle|\underline{0}\rangle + a_1|1\rangle|\underline{0}\rangle\right). \tag{9}$$

Otherwise, when the outcome is 1 we have

$$\frac{1}{2}\left(a_0|0\rangle|\underline{1}\rangle - a_1|1\rangle|\underline{1}\rangle\right). \tag{10}$$

To recover a deterministic pure state, a controlled $Z$ gate is applied to (10), correcting the phase flip and obtaining (9). For a larger GHZ state, this correction involves a $Z$ gate controlled by the mod-2 sum of all measured qubits instead.

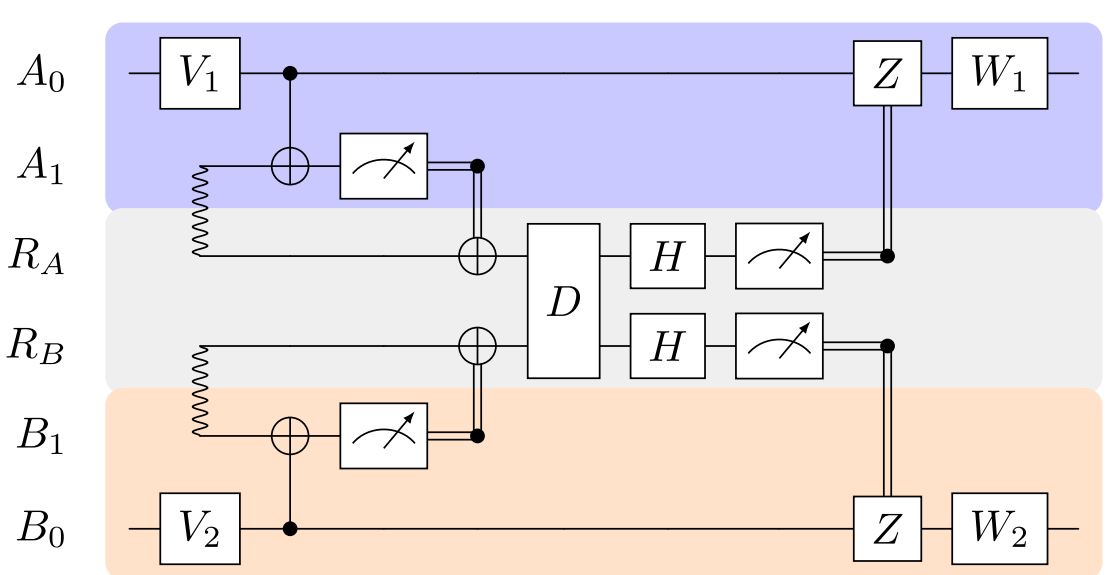

**FIGURE 5.** Using cat-entangler and cat-disentangler protocols with a central node, a two-qubit gate $U$ between qubits $R_A$ and $R_B$ can be applied remotely. Here, $U$ is decomposed into a two-qubit gate $D$, diagonal in the computational basis of $R_A$ and $R_B$, and local operations $V_i$ and $W_i$, as $U = (V_1 \otimes V_2)D(W_1 \otimes W_2)$.

## III. RESULTS

In this article, we study DQC systems with a central node that carries out all of the entangling operations with the remaining nodes. This central node acts as a router, handling Bell pair creation, asynchronous measurements, and classical information transfer. However, it also participates in the computation of distributed quantum operations.

Let us begin with two states $|\psi\rangle = a_0|0\rangle + a_1|1\rangle$ and $|\varphi\rangle = b_0|0\rangle + b_1|1\rangle$. After expanding each state by a cat-entangler operation with a Bell pair shared with the same central node, the resulting state becomes

$$\begin{aligned} &a_0 b_0 |00\rangle|00\rangle + a_0 b_1 |01\rangle|01\rangle \\ &\quad + a_1 b_0 |10\rangle|10\rangle + a_1 b_1 |11\rangle|11\rangle \end{aligned} \tag{11}$$

which enables local operations in the qubits of the central node (the second register), as shown in Fig. 5. The operations in the router must all be diagonal on the basis of the fan-out qubits, limiting their action to phase shifts.

Many gates in relevant quantum algorithms can be locally diagonalized, making this constraint less restrictive. Some common examples are multicontrolled gates, including the Toffoli gate (for three or more qubits) or multiqubit versions

**Algorithm 1:** Pseudocode for distributed MCZ gate with central node among $k$ QPUs.

**Require:** QPUs $q \in \{q_0, \ldots, q_k\}$ have link to central node
**Require:** $q$ has qubits $q[0]$ for communication and $q[1, \ldots, n_q]$ for data
**Require:** Central node $R$ has qubits $R[q]$ $\forall q \in \{q_0, \ldots, q_k\}$

**Step 1: Generate ebits**
1: **for** q $\in \{q_0, \ldots, q_k\}$ **in parallel do**
2: **for** retry $\in \{1, \ldots, \text{max_retries}\}$ **do**
3: attempt[$q$] $\leftarrow$ GenerateEbit($q[0]$, $R[q]$)
4: **if** attempt[$q$] $=$ True **then**
5: **break** ▷ Exit loop
6: **else**
7: Wait(delay) ▷ Delay before retry
8: **if** not all(attempt $=$ True) **then**
9: **return** Fail ▷ Program failed

**Step 2: Apply local operations on QPUs**
10: **for** $q \in \{q_0, \ldots, q_k\}$ **in parallel do**
11: MCZ($q[0], \ldots, q[n_q]$) ▷ Local gate on QPU $q$
12: $m_q \leftarrow \text{Measure}_X(q[0])$
13: Send $m_q$ to central node ▷ Classical bit

**Step 3: Apply local operations on central node**
14: **for** $q \in \{q_0, \ldots, q_k\}$ **do**
15: H($R[q]$) ▷ Change basis
16: **if** $m_q = 1$ **then** ▷ Probability 1/2
17: X($R[q]$) ▷ Correction
18: MCZ($R[q_0], \ldots, R[q_k]$) ▷ Gate on central node
19: **for** $q \in \{q_0, \ldots, q_k\}$ **do**
20: $m_q \leftarrow \text{Measure}_X(R[q])$
21: Send $m_q$ to QPU $q$ ▷ Classical bit

**Step 4: Apply local corrections on QPUs**
22: **for** $q \in \{q_0, \ldots, q_k\}$ **in parallel do**
23: **if** $m_q = 1$ **then** ▷ Probability 1/2
24: MCZ($q[1], \ldots, q[n_q]$) ▷ Correction
25: **return** Success

of the $R_{ZZ}$ gate. The original Eisert–Jacobs–Papadopoulos–Plenio (EJPP) telegate protocol [46] and several more recent works [45], [62] describe how to teleport multiqubit gates between many parties. In this work, we focus on the EJPP protocol, which was proven to be optimal in the required resources, and work seamlessly with the star network structure and the cat-entangler/disentangler formalism.

### A. COLLECTIVE OPERATIONS USING A CENTRAL NODE: DISTRIBUTED MCZ GATE

We will now discuss how collective MCZ gates can be distributed among $k$ nodes, following a similar scheme as the EJPP protocol, and using a star network architecture with a central node as an active computational element. The distribution of the $N$-qubit MCZ gate among $k$ nodes can be decomposed into three steps: first, a $(\lceil \frac{N}{k} \rceil + 1)$-qubit MCZ gate is applied in each node,[1] which also acts on the communication qubit. After this operation, the communication qubit is measured, and the resulting classical bit is sent to the corresponding router qubit, producing the cat-state (fan-out). Second, a $k$-qubit MCZ gate in the central node operates between the fan-out qubits. Finally, the router qubits are measured to complete the cat-disentangler (fan-in), and the resulting classical bits are used to control a $\lceil \frac{N}{k} \rceil$-qubit MCZ gate acting on the data qubits of each node. Notably, these last MCZ operations are conditional and are only executed half of the time on average. The step-by-step pseudocode to implement a distributed MCZ gate is compiled in Algorithm 1, and the equivalent quantum circuit for three nodes following this protocol is shown in Fig. 6.

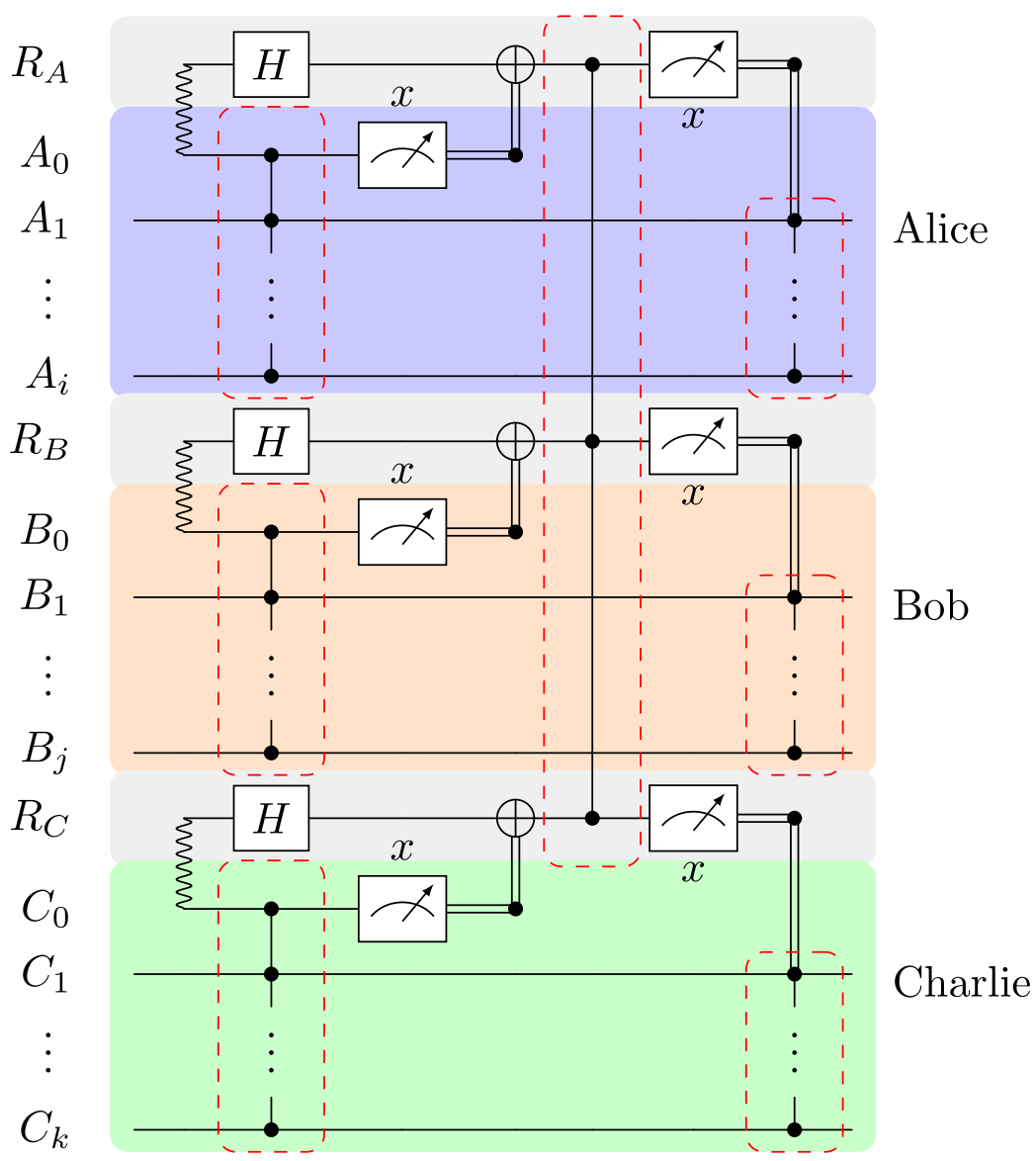

**FIGURE 6.** The simplest collective gate that can be used to showcase the utility of this architecture is an MCZ gate distributed across a network with multiple nodes connected by a router. The router $R$ requires a single ebit per node, and no direct quantum or classical communication is required between workers. In the picture, an $N = i + j + k$-qubit nonlocal gate is implemented with registers of $i$, $j$, $k$ data qubits in nodes Alice, Bob, and Charlie, respectively, and three qubits in $R$. The 0-indexed qubits in each node act as communication qubits. The multicontrolled gates (outlined in red dashed lines) are local operations in their respective devices and could be implemented directly or decomposed in one- and two-qubit gates. Notice how the last $X$-basis measurements control smaller MCZ gates, which are not required when the outcome of their respective measurements is 0 (i.e., 50% of the time).

### B. GENERAL DISTRIBUTED DIAGONAL GATES

For a more general case, we will now show that any diagonal gate can be teleported using this protocol. For simplicity, let us begin with a separable state of three qubits, belonging to Alice, Bob, and Charlie, respectively, even though this method is generalizable to $N$ qubits. The joint state of these three qubits can then be written as

$$|abc\rangle = \sum_{i,j,k=\{0,1\}} a_i b_j c_k \, |ijk\rangle \,. \tag{12}$$

[1] Assuming equal size nodes for simplicity. When $N$ is not divisible by $k$, one or more nodes utilize fewer qubits.

**FIGURE 7.** Gate lumping in a single teleportation round. Multiple diagonal gates can be consecutively performed in the router, at no additional cost in the number of ebits required [60], [61]. (a) Some relevant gates that can be executed in the router in one round of fan-out and fan-in, from two-qubit $CR_z(\theta)$ or $R_{zz}(\theta)$ gates, to larger $CCR_z(\theta)$ and $R_{zzz}(\theta)$ gates. In (b) and (c), the blue/red lines represent local/nonlocal interactions. (b) Communication-dense algorithms using regular teledata or telegate grow quadratically with the number of nodes and communication qubits per node if no router is used. (c) Using the router as an active computing node, the ebit cost becomes linear. When all of the gates can be locally diagonalized, they can be executed in a single round of teleportation. Notice that to produce the Bell pairs between nodes in (b), the number of ebits required either scales quadratically with the number of qubits per node $n$ and the number of nodes $k$ as $O(n^2k^2)$ or a separate node would be required to distribute entanglement, increasing the number of ebits required even further.

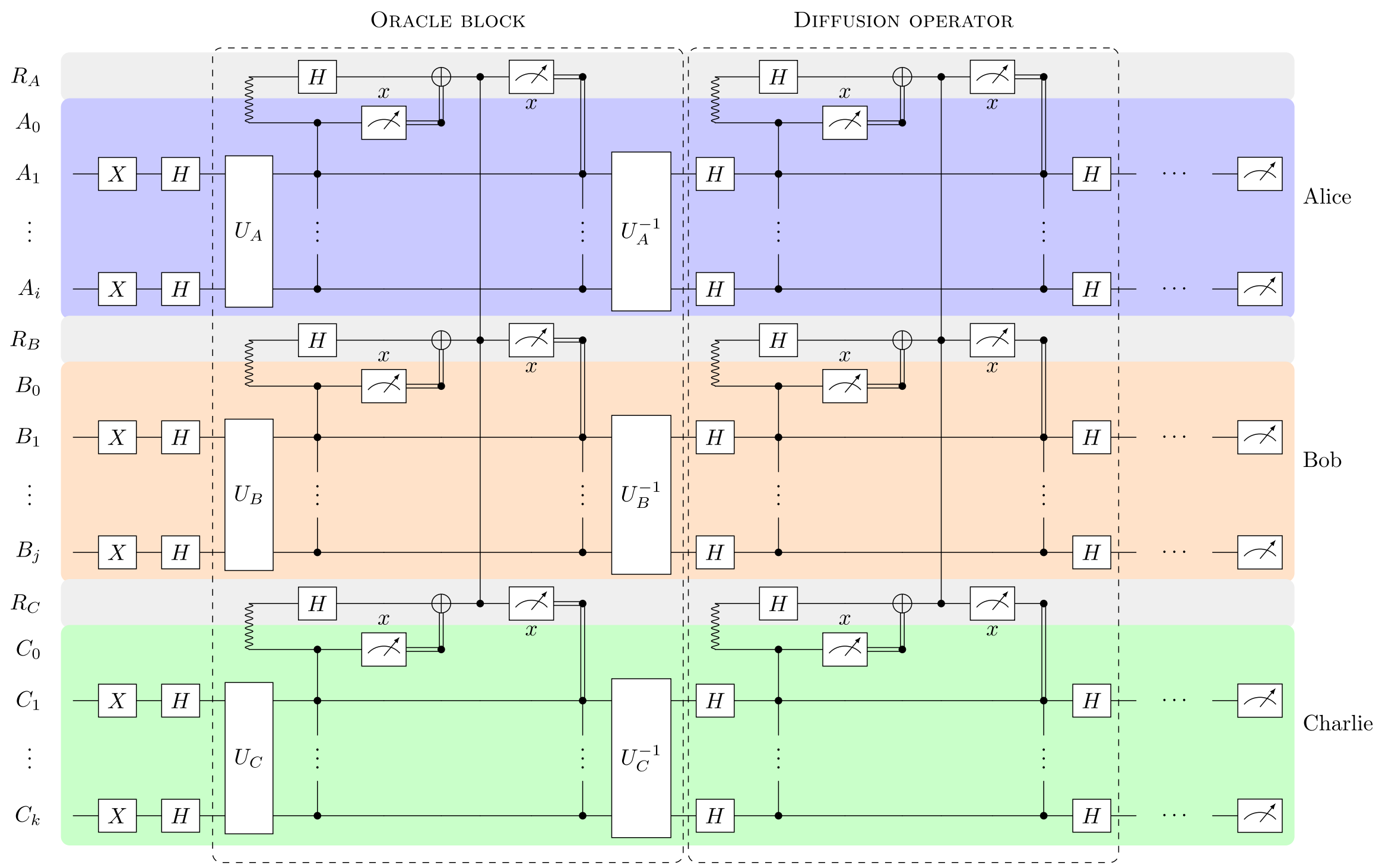

**FIGURE 8.** Distributed Grover implementation, showcasing indirect connectivity between QPUs, with quantum links only between them and the router. Each Grover block requires two collective MCZ operations, which can be implemented using the techniques proposed in this article. Here, the $U_i$ gates transform the MCZ into the desired oracle locally, which marks a subset of the search space with a phase flip. In the simplest case, a classical bitstring is encoded by applying $X$ gates in selected qubits. Although general implementations of Grover's algorithm may require additional connectivity between nodes, the total number of collective operations grows linearly with the number of Grover blocks, $O(\sqrt{N})$, where $N$ is the size of the search space.

Applying fan-out to each of the qubits, the state becomes

$$|ABC\rangle = \sum_{i,j,k=\{0,1\}} a_i b_j c_k \, |ijk\rangle \, |ijk\rangle \, . \tag{13}$$

The second register of the state corresponds to the router qubits. Applying a general diagonal unitary $\mathbf{M}$, parameterized by its eigenvalues $e^{i\theta_{lmn}}$

$$\mathbf{M} = \sum_{l,m,n=\{0,1\}} e^{i\theta_{lmn}} |lmn\rangle\langle lmn| \tag{14}$$

over the router qubits, corresponds to applying $(\mathbb{1} \otimes \mathbf{M})$ $|ABC\rangle$ to the whole expanded state, obtaining

$$\sum_{i,j,k} a_i b_j c_k \, |ijk\rangle \sum_{l,m,n} e^{i\theta_{lmn}} \, |lmn\rangle \, \delta_{ijk,lmn}$$
$$= \sum_{i,j,k} a_i b_j c_k e^{i\theta_{ijk}} \, |ijk\rangle \, |ijk\rangle \tag{15}$$

where $\delta_{ijk,lmn}$ is the Kronecker delta. The result of (15) is equivalent to applying phase shifts $e^{i\theta_{ijk}}$ on the second register, translated to the first register by phase-kickback. The algorithm ends with a fan-in operation

$$\sum_{i,j,k} e^{i\theta_{ijk}} a_i b_j c_k \left|ijk\right\rangle \left|ijk\right\rangle \xrightarrow{\text{fan-in}} \sum_{i,j,k} e^{i\theta_{ijk}} a_i b_j c_k \left|ijk\right\rangle \quad (16)$$

so that the remote action of the unitary is teleported back to the remote qubits in Alice, Bob, and Charlie's registers.

Diagonal gates can be constructed by consecutive application of $CR_z(\theta)$, $CCR_z(\theta)$, $R_{ZZ}(\theta)$, and $R_{ZZZ}(\theta)$ gates, to name a few, as shown in Fig. 7. Circuits, such as the QAOA ansatzes, often include sequences of $R_{ZZ}(\theta)$ gates between the circuit qubits following some coupling graph. As all these gates commute, they can be executed in one single round of teleportation, although several Bell pairs for each node may be necessary depending on the graph topology. This also extends to larger multiqubit rotation gates [63]. For a general uninformed case, a diagonal gate acting over $N$ qubits can be written using $O(2^N)$ CNOT gates [64] plus single-qubit rotations. Teleporting each of these CNOT gates would become unfeasible. However, with this procedure, these gates can be applied locally in the router qubits. So, any diagonal gate can be teleported with a minimum of $k$ and a maximum of $N$ ebits.

### *C. DISTRIBUTED GROVER ALGORITHM*

We will now describe how these tools can be applied to Grover's unstructured search algorithm. As the aim of this section is purely illustrative, a textbook implementation of Grover's algorithm will be used. In this simple case, the algorithm is used for finding one specific binary string of $N$ bits. The monolithic circuit consists of a succession of $N$-qubit MCZ gates, plus single-qubit gates, such as $H$ and $X$ gates. Each layer of Grover's algorithm includes two such MCZ gates—one for the oracle and another for the diffuser operators—and the number of layers that maximizes the probability of obtaining the correct solution grows as $O(\sqrt{2^N})$ [26]. While general Grover oracles for more realistic problems can be more complex, we expect a large class of oracles to be described using a diagonal plus local decomposition.

The quantum gate circuit of this implementation is shown in Fig. 8. The MCZ gates are distributed across $k$ nodes, with a router of at least $k$ qubits. In order to accommodate the whole MCZ gate, the nodes must have a minimum of $\lceil \frac{N}{k} \rceil$ data qubits each (if all nodes are equal), plus one communication qubit that will share Bell pairs with the router. Compared to the monolithic case, and assuming that communication qubits can be reused, a constant overhead of $2k$ additional qubits is required for communication ($k$ in the router plus one in each of the $k$ nodes), not depending on the number of layers or MCZ gates to be distributed, and the ebit count is $2k$ per Grover layer. Since the writing of this article, we have become aware of a recent preprint by Feng et al. [65], where the communication complexity of several distributed algorithms is discussed. A linear scaling is found for Grover's algorithm, which is consistent with our results. The code for our implementation can be found in the Zenodo repository.[2]

**Algorithm 2:** Pseudocode for QAOA($\gamma, \beta$) layer with central node among $k$ QPUs for $n$-qubit MaxCut.

**Require:** QPUs $q \in \{q_0, \ldots, q_k\}$ have link to central node
**Require:** $q$ has qubits $q^c[0, \ldots, c_q]$ for communication and $q[0, \ldots, n_q]$ for data ($n_q \geq c_q$)
**Require:** Central node $R$ has qubits $R[q][0, \ldots, c_q]$ $\forall q \in \{q_0, \ldots, q_k\}$
**Step 1: Generate ebits (see Algorithm 1)**
**Step 2: Apply local operations on QPUs**
1: **for** $q \in \{q_0, \ldots, q_k\}$ **in parallel do**
2: **for** $i \in \{0, n_q\}$ **do**
3: $H(q[i])$ ▷ Init $|+\rangle^{\otimes n}$ state (only 1st layer)
4: **for all** $i \in \{0, n_q\}$, $j \neq i \in \{0, n_q\}$ **do**
5: $R_{zz}(\gamma, q_i, q_j)$ ▷ Cost layer
6: **for** $i \in \{0, \ldots, c_q\}$ **do**
7: CZ($q[i], q^c[i]$) ▷ Fan-out border qubits
8: $m_q[i] \leftarrow \text{Measure}_X(q^c[i])$
9: Send $m_q[i]$ to central node ▷ Classical bit
**Step 3: Apply local operations on central node**
10: **for** $q_l \in \{q_0, \ldots, q_k\}$ **do**
11: **for** $i \in \{0, \ldots, c_{q_l}\}$ **do**
12: H($R[q_l][i]$) ▷ Change basis
13: **if** $m_{q_l}[i] = 1$ **then**
14: X($R[q_l][i]$) ▷ Correction
15: **for all** $q_l \in \{q_0, \ldots, q_k\}$, $q_m \in \{q_{i+1}, \ldots, q_k\}$ **do**
16: **for all** $i \in \{0, \ldots, c_{q_l}\}$, $j \in \{0, \ldots, c_{q_m}\}$ **do**
17: $R_{zz}(\gamma, R[q_l][i], R[q_m][j])$ ▷ Cost layer
18: **for** $q_l \in \{q_0, \ldots, q_k\}$ **do**
19: $m_{q_l}[i] \leftarrow \text{Measure}_X(R[q_l][i])$
20: Send $m_{q_l}[i]$ to QPU $q_l$ ▷ Classical bit
**Step 4: Apply local corrections on QPUs**
21: **for** $q \in \{q_0, \ldots, q_k\}$ **in parallel do**
22: **for** $i \in \{0, \ldots, c_q\}$ **do**
23: **if** $m_q[l] = 1$ **then** ▷ Probability 1/2
24: $Z(q[i])$ ▷ Correction
25: **for** $i \in \{0, \ldots, n_q\}$ **do**
26: $R_X(\beta, q[i])$ ▷ Mixer layer
27: **return** Success

### *D. APPLICATION TO QAOA*

To complement the Grover case study, we apply the proposed techniques to the QAOA for the MaxCut problem. We consider distributed star-like architectures where several QPUs connect to a central node. Each QPU is modeled with all-to-all local connectivity between its qubits and a set of communication qubits linking to the router. For this purpose, we simulate one QAOA layer over a grid of parameters $\{\beta, \gamma\}$, performing 100 shots for each pair, and calculate the MaxCut value from the resulting counts. For comparison, we implement both monolithic and distributed versions. Algorithm 2 compiles the logic for network distribution of this protocol.

[2]https://doi.org/10.5281/zenodo.17376198

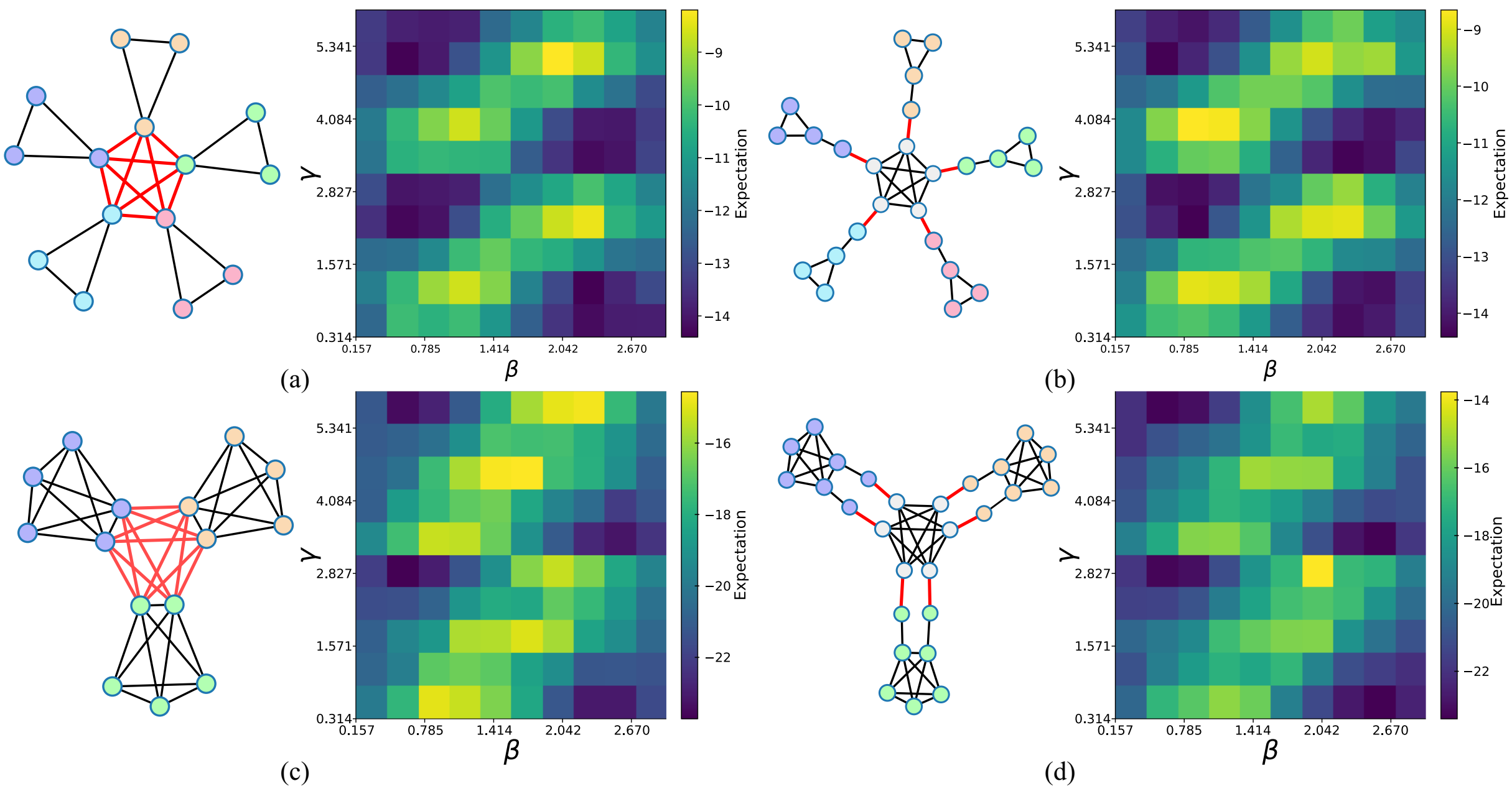

**FIGURE 9. Optimization landscapes of single-layer QAOA for a 15-qubit MaxCut problem, comparing monolithic and distributed executions using our cat-assisted router protocol. Simulations were carried out using Qiskit Aer. The MaxCut value is shown as a function of the variational parameters $\beta$ and $\gamma$ (100 shots per point, $\beta$ spans $(0, \pi)$ and $\gamma$ spans $(0, 2\pi)$). Black lines indicate local $R_{zz}(\gamma)$ rotations while red lines indicate inter-QPU links. (a) Monolithic problem directly split in five three-qubit QPUs. (b) Five four-qubit QPUs plus five-qubit central node. (c) Monolithic problem with 15 qubits split into three five-qubit QPUs. (d) Three five-qubit + two-communication-qubit QPUs plus a six-qubit router, reducing the number of inter-QPU connections from 12 to 6. In both scenarios, the optimization landscapes are identical up to shot noise, demonstrating that the distributed protocol faithfully reproduces the monolithic performance while lowering communication overhead. (a) Monolithic 5 × 3 qubits. (b) Distributed 5 × 4 + 5 qubits. (c) Monolithic 3 × 5 qubits. (d) Distributed 3 × 7 + 6 qubits.**

Two representative examples are shown in Fig. 9. Both represent a 15-qubit MaxCut instance, with different topologies. In the first example, five three-qubit clusters form a monolithic 15-qubit instance. The distributed protocol instead uses five four-qubit QPUs (one qubit for communication) and a five-qubit router, so while the total number of qubits is larger, each individual device is smaller and easier to realize in practice. In the swap-and-teleport scenario, ten teleports (20 ebits if routing through entanglement swapping) would be required, and as the number of cross-QPU gates exceeds the available communication qubits, several nonlocal rounds would be necessary. However, our protocol completes the layer in a single round with five ebits. In the second example, we use three five-qubit clusters and two connections to every other cluster. The distributed version uses three seven-qubit QPUs (two qubits for communication) plus a six-qubit router. A swap-and-teleport approach would require 12 teleports (24 Bell pairs if routed). In all tested cases, the optimization landscapes obtained in the monolithic and distributed executions are indistinguishable (up to shot noise variance), confirming that our distributed protocol reproduces the same variational performance as the monolithic implementation.

For a more general number of qubits, while still preserving this star-like network topology, the number of cross-QPU operations would grow as $O(k^2M^2)$, for $k$ QPUs and $M$ communication qubits per QPU. Our router-assisted method scales only linearly as $kM$, as each QPU connects only to the central router. These results illustrate that the router-assisted method preserves algorithmic fidelity while reducing both ebit cost and nonlocal execution time across different distributed layouts.

## IV. DISCUSSION

This article proposes a resource-efficient DQC framework, combining the EJPP telegate protocol [46] with a star network structure. In this section, we will discuss how this framework addresses current challenges in DQC, such as applicability and scalability. We will also discuss how partitioning circuits this way can offer a circuit depth and, therefore, runtime, reduction. Finally, we will elaborate on how these techniques can be extended to other algorithms, advancements, and network structures.

### A. APPLICABILITY AND SCALABILITY

Practical applicability comes from adopting the benefits of a network structure, offloading most of the network-dependent workload to the central node. Therefore, the nonlocal gate execution and entanglement generation of each computing node can be limited from the computing nodes to the central node, unaware of the actions of the remaining nodes. The central node can schedule and parallelize the individual cat-entangler and disentangler protocols with each individual node as needed. Another advantage of this network structure is its reconfigurability, as nodes can join or leave the network without disturbing the remaining nodes by simply establishing a new connection with the central node. Using the MCZ, $R_{ZZ}$, and single-qubit gates as building blocks, many useful gates can be generated. For instance, the $n$-Toffoli gate can be obtained by applying $H$ gates before and after the target qubit. Anticontrols can also be obtained by applying $X$ gates before and after the controls. Any other multiqubit rotation gate, such as $R_{XY}(\theta)$, can be diagonalized to the computational basis with single-qubit gates, such as $H$ and $S$ gates.

Moreover, scalability comes from the ability to handle operations that exceed the capacity of individual QPUs, such as multiqubit Toffoli gates. The optimal cost of $k$ ebits for $k$ nodes in this architecture scales better than the $2k$ required when following the entanglement swapping model and is guaranteed from the optimality of the EJPP telegate protocol. For two-qubit operations, even though the ebit cost advantage is lost, the use of the central router is equivalent to that of entanglement swapping.

However, even for two-node operations, the star network offers a potential improvement in communication scaling by reducing the number of required ebits in certain, strongly entangled scenarios. To understand this, consider a system of $k$ nodes, each with $M_i$ nonlocal gate executions with the rest. To avoid ambiguity, here we define a *nonlocal gate Singular: execution* as a required multiqubit gate between qubits located in different QPUs. For example, if node A hosts qubits $q_1$ and $q_2$ and node B hosts qubits $q_3$ and $q_4$, a circuit that included gates $CZ(q_1, q_3)$ and $CZ(q_2, q_4)$ would have two nonlocal gate executions between A and B.

In a fully connected network, the number of pairwise interactions, i.e., nonlocal gates, grows as $\sum_{i>j}^{k} M_i M_j$, where $i$ and $j$ are index nodes (i.e., the product of cross-qubit counts between all node pairs). Each of these nonlocal gates requires teleporting its own ebit and classical bit. For simplicity, considering $M_i = M$ for all QPUs, the number of interactions grows as $\binom{k}{2} M^2$. As an illustration, in a four-node system where each node is connected to all others, the total number of interactions is the sum over all cross-partition two-qubit gates. If each pair of nodes exchanges two CZ gates, then the system has $2 \times \binom{4}{2} = 12$ interactions. In contrast, the star network topology significantly reduces this cost, as shown in Fig. 7(b) and (c). In this topology, all nonlocal gate executions are mediated through a central router. Instead of requiring direct pairwise teleportation between nodes, each node simply requires generating entangled pairs with the central router, and the router facilitates the necessary entanglement for communication across the network. For diagonal operations, such as those used in this protocol, the router only needs to establish $M$ ebits for each node, resulting in a total of $kM$ ebits and classical bits. The advantage becomes clearly apparent for circuits where many qubits in each node need to communicate with many qubits in the rest of the nodes.

### B. EXECUTION TIME ESTIMATES

To complement the resource analysis, in this section, we provide an analytic execution time calculation that distinguishes between local and nonlocal contributions to circuit depth, considering compilation with a one- and two-qubit native gate set. In distributed settings, the dominant cost typically comes from the creation of Bell pairs and classical message passing. The network scheme presented in this article is agnostic to the specific values of these times, relying only on the general assumption that

$$\tau_{\text{EPR}} + \tau_{\text{classical}} \gg \tau_1, \tau_2 \tag{17}$$

where $\tau_{\text{EPR}}$ is the expected time to obtain a Bell pair, $\tau_{\text{classical}}$ is the latency of a classical message round, and $\tau_1$ and $\tau_2$ denote single- and two-qubit gate times, respectively.

We define the nonlocal depth $D_{\text{nonlocal}}$ as the number of sequential communication rounds needed to execute all cross-QPU gates. In a swap-and-teleport baseline, $D_{\text{nonlocal}}$ is lower bounded by the edge-coloring number of the cross-QPU gate graph: if a qubit participates in $r$ gates with distinct partners, then these must be scheduled in $r$ separate nonlocal rounds. In the proposed scheme, all such gates can be performed in a single nonlocal round per batch, while leaving some additional local depth (two-qubit gates within each QPU and diagonal operations at the router). As a result, the total execution time for a nonlocal batch of an MCZ gate can be expressed as

$$T_{\text{swap}} \approx D_{\text{nonlocal}} \cdot (\tau_{\text{EPR}} + \tau_{\text{classical}}) + \max_i D_i \tau_2 \tag{18}$$

$$T_{\text{cat}} \approx (\tau_{\text{EPR}} + \tau_{\text{classical}}) + 2 \max_i D_i \tau_2 + D_{\text{R}} \tau_2 \tag{19}$$

where $T_{\text{swap}}$ and $T_{\text{cat}}$ are the swap-and-teleport and this work's execution times, respectively, $D_i$ is the local depth at QPU $i$ after compilation, and $D_{\text{R}}$ is the local depth at the router. In Section IV-C, we connect these expressions with CX-depth reductions in MCZ and Grover circuits.

This analysis shows that our protocol is agnostic to the physical values of the device parameters: it consistently reduces the nonlocal depth of a batch from $D_{\text{nonlocal}}$ rounds to one, as long as all operations performed in the same batch are diagonal. This scheme trades expensive nonlocal operations with faster local QPU gates, always under the assumption in (17). The relative advantage grows with cross-QPU connectivity: when each qubit must interact with many partners, swap-and-teleport requires multiple nonlocal rounds, while our method reduces them to a single round. Examples of this, presented in this article, are wide multicontrolled operations, or QAOA cost layers on dense graphs.

### C. CIRCUIT DEPTH REDUCTION

As a side effect of partitioning operations across $k$ nodes, the distribution of large collective gates can offer a circuit depth reduction in certain cases, e.g., when the native gate set of the nodes is limited to one- and two-qubit gates. Given their prevalence in quantum algorithms, many decompositions are known for the $N$-qubit Toffoli gate, equivalent up to single-qubit gates to the MCZ gate used in this article, into CNOT+T, with or without ancillae. Using Nielsen and Chuang's [26] $6N - 12$ estimate for CNOT gate depth for an $N$-qubit Toffoli gate, running large MCZ distributed gates following the technique proposed yields a depth scaling $O(\frac{12N}{k} + 6k)$: as the gates are broken into multiple chunks that are run in different nodes, these pieces can be run in parallel. Nevertheless, note that a $k$-qubit gate must be executed in the router, which will bottleneck this improvement for $k \to N$, where $k = \sqrt{2N}$ being the optimal number of nodes up to the rounding effect. As shown in Fig. 10(b), this optimal $k$ results in a speedup of $O(\sqrt{N})$, a behavior that can be found using any other linear scaling decomposition. The previous argument can be applied to Grover's algorithm in this article. While other decompositions of the $N$-qubit Toffoli gate into larger unitaries exist, such as regular three-qubit Toffoli gates ($\sim 16N$ gates) [66] or GMS gates in the case of trapped ions ($\sim 3N$ gates) [67], these also scale linearly, so these arguments stand. As shown in Section IV-B, the CX-depth reduction translates into a proportional runtime reduction, as each extra depth layer corresponds to a nonlocal round of duration $\tau_{\text{EPR}} + \tau_{\text{classical}}$. Thus, Fig. 10(b) not only shows circuit depth parallelization but also predicts runtime speedup under realistic assumptions.

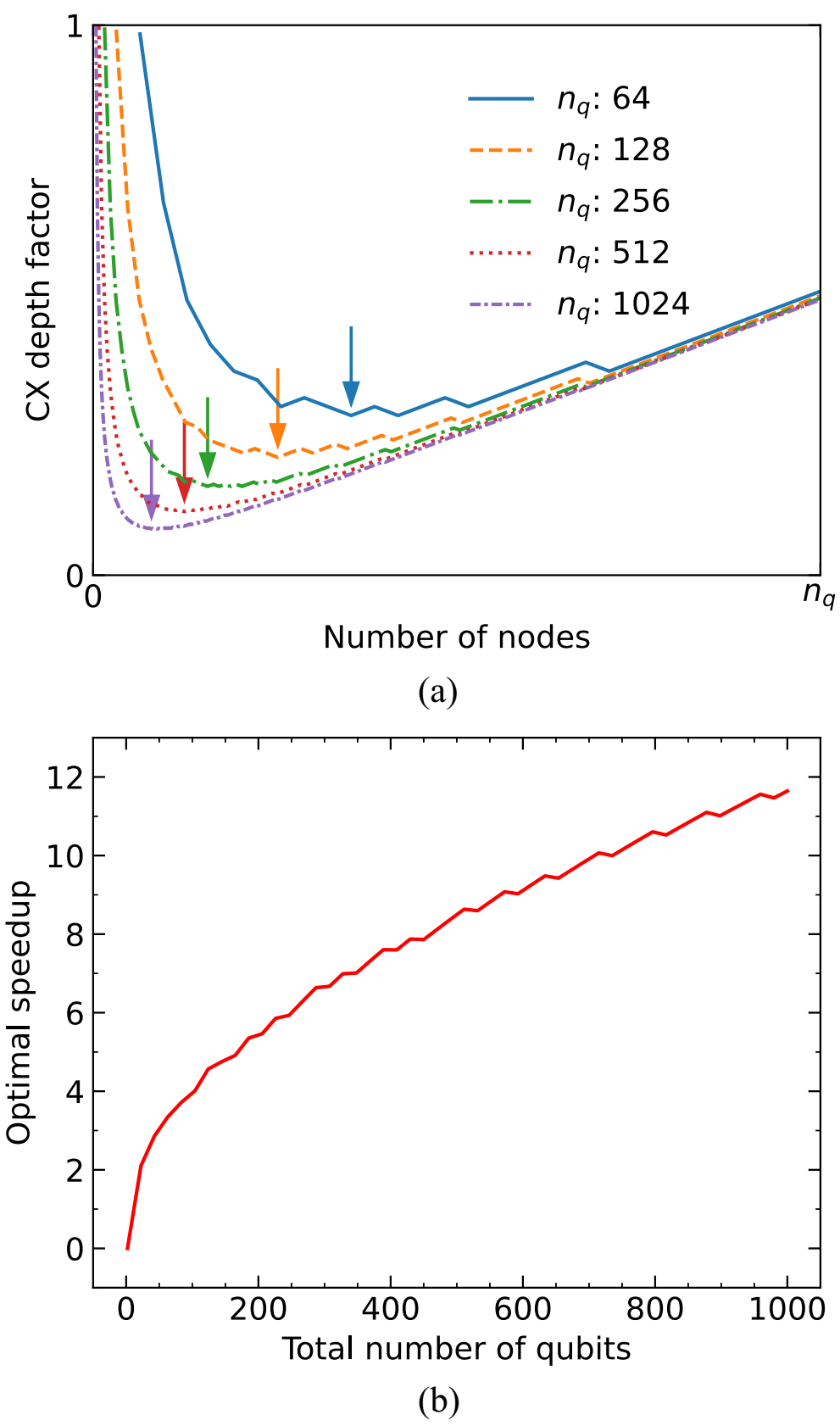

**FIGURE 10. Scaling of the distributed MCZ protocol. (a) CX depth of the distributed MCZ decomposition divided by the depth of the monolithic circuit, as a function of the number of nodes, following Nielsen and Chuang's estimation [26]. Partitioning the circuit in a number of nodes rapidly decreases the circuit depth up to a saturation point, above which the depth of the circuit in the central node dominates. The optimal value is found at $\sqrt{2N}$ (marked by the arrows). (b) Optimal speedup (inverse of the optimal depth factor) as a function of the total number of qubits (excluding the router), growing as $O(\sqrt{N})$ up to a rounding.**

While we are aware of recent, more efficient decompositions that may achieve sublinear scaling in circuit depth [68], [69], these include either a growing number of ancillae, large prefactors, or approximations. We would expect that for sublinear depth scaling, monolithic computation of MCZ gates may be better for a large number of qubits.

Even though the total number of gates to be executed in the whole network is larger than that of the monolithic execution, parallelization across the $k$ nodes mitigates this overhead. The total ebit cost of this computation is $k$, although Bell pairs could be produced and consumed simultaneously if the router technology allows it. In addition, the classical control means that the $\lceil \frac{N}{k} \rceil$-qubit MCZ gates appear only half of the time, reducing gate counts. Nevertheless, this has no effect on the depth estimation for large $k$—any depth reduction would require all measured bits to be 0, an outcome that becomes exponentially unlikely as $k$ increases.

### D. RELATION TO OTHER WORK

This technique can be applied to other algorithms in the literature that include multicontrolled gates. To give some recent examples, the maximum independent set problem is solved in [70] with a constrained version of QAOA, including a mixing unitary built from $n$-Toffoli gates. Both the $R_{ZZ}(\theta)$ part of the ansatz and this larger mixing unitary can be efficiently distributed using our MCZ protocol. In [71], a distributed quantum phase estimation is proposed, including a reflection operator similar to Grover's diffuser. To distribute this operator, they first decompose it into several two-qubit gates, which are then teleported across the nodes. Remarkably, using the distributed MCZ protocol in our work, the cost of this operation would be reduced to a single ebit per node. A distributed SAT solver that uses a parallel version of the oracle and diffuser operators was also introduced in [72], significantly reducing the depth of the circuit. However, it still requires distributing a collective multicontrolled $n$-Toffoli gate, which could be efficiently handled by our distributed MCZ protocol.

Other improvements for saving ebits in the literature should be applicable to this protocol, e.g., in [60] and [61], the EJPP protocol [46] is extended by merging nonsequential distributing processes in a "packing method" or "embedding" to save ebits. Wu et al. [60] considered circuits consisting of only one- and two-qubit gates, such as ansatzes for variational algorithms, distributed between two nodes. Afterward, their packing/embedding method has been extended to multipartite scenarios in [61], distributing circuits across homogeneous and heterogeneous networks. A combination of these methods and collective operations in the star topology could further improve the ebit counts for DQC for many algorithms.

Furthermore, when accounting for Bell pair creation rates and communication latency, it is important to carefully consider the tradeoffs between telegate and teledata approaches. Van Meter et al. [29] found that teledata combined with a linear topology outperformed telegate for specific tasks, such as carry-ripple adders. While their findings favor a topology and communication strategy different from the one proposed in this work, the conclusions are highly dependent on the structure and communication patterns of the algorithms studied. Our approach, focused on collective multiqubit operations in star-like topologies, addresses a different class of problems, for which centralized coordination and minimal ebit consumption per gate can offer practical advantages.

The benefits of a star topology have already been discussed for the distribution of multipartite entanglement [73], [74]. While our work does not rely on multipartite entanglement as a resource but instead on individual Bell pairs between the central nodes and each computing node, several considerations on the ebit fidelity, noisy teleportations, and network latencies remain applicable. The study of realistic conditions for DQC in real hardware or DQC emulators could provide insight into the applicability and scalability of these techniques in the short-medium term and the feasibility of running DQC algorithms on noisy devices with hardware-specific optimizations, which could improve this protocol for specific technologies.

Our proposal introduces additional communication overhead and requires the involvement of a central router, although this can be justified when part of the computation is clusterized. In the star architecture, the central router can become a bottleneck, limiting the scalability if the bandwidth of the router (e.g., Bell pair creation and number of qubits available) is unable to meet the demands of the network. Another issue with star networks is that a central router becomes a potential single point of failure.

Therefore, future research should investigate the extension of this protocol to higher tier networks, commonly used in HPC environments, to solve these issues. The star topology presented here could be generalized into a hierarchy of routers, where each routing tier acts as a central node for the tier below [see Fig. 2(e)]. In such networks, redundant routing elements can enhance robustness, providing multiple possible communication paths. Extending our scheme to this setting would require upper tier routers to coordinate fan-out operations across subnetworks, increasing overhead; this is analogous to the performance difference of intra/internode and interrack communication of MPI in classical HPC systems, where not all communication links have equal computational cost or latency [9], [75].

Finally, further work should focus on extending the scope of this work to include efficient teleportation of general multiqubit gates and tightening the bounds for the ebit cost of these operations. As already indicated in the original EJPP paper, constructing an optimal procedure for general quantum gates is not trivial, but specific gate decompositions can be found that take advantage of these techniques [46].

## V. CONCLUSION

In this article, we have introduced a network-assisted scheme for collective operations in DQC, where a central router actively participates in both communication and computation. Using nonlocal fan-out operations, we have demonstrated that collective operations can be performed with the aid of quantum routers, resulting in significant ebit and execution time savings. These ebit savings begin in the case of multicontrolled quantum operations and can grow large when operations can be lumped, i.e., commuting operations that can be distributed simultaneously in a single round of teleportation [60], [61]. With two-node operation, the ebit cost is equivalent to the telegate protocol [46] using entanglement swapping and scales better in the case of operations that require more than two nodes. Therefore, we show that entanglement swapping, while very tailored toward quantum Internet applications, may be suboptimal for certain DQC algorithms. An analytic execution time estimate complements the resource analysis, separating nonlocal communication rounds from local two-qubit gate depth, and clarifies under which conditions centralization offers a runtime advantage.

This strategy is particularly beneficial for algorithms with many collective gates. We showcase that Grover's search can be implemented more efficiently using this method than with swap-and-teleport methods. In these settings, centralization effectively trades multiple rounds of entanglement generation and communication for inexpensive local operations.

Beyond Grover's search algorithm, our application to single-layer QAOA shows that the monolithic and distributed implementations yield identical optimization landscapes, confirming algorithmic fidelity. These toy models have star-like partitioning, chosen to illustrate correctness and resource savings in a controlled setting. In practice, QAOA graphs often are unstructured, so the advantages of our method will depend on both the problem graph, and how it is partitioned across QPUs: certain clusterizations may saturate connectivity and provide little to no advantage, while others can significantly reduce nonlocal depth. This mirrors classical HPC practice, where the benefit of collective operations depends on where communication occurs (within a node, within a rack, or across racks). We leave the study of realistic partitioning strategies for future work.

By introducing quantum routers as active computational nodes, this work paves the way for resource-efficient and scalable quantum architectures. Future advancements in hardware implementation and multirouter extensions could further enhance the applicability of this framework to a broader range of algorithms and network topologies. As quantum technologies continue to evolve, the integration of network-assisted computation stands out as a promising pathway toward large-scale, fault-tolerant quantum systems.

While the present results are analytical, validation in realistic network conditions, including error models, different topologies, and algorithms beyond Grover and QAOA, remains as future work. Overall, our findings show that delegating both communication and computation to a central device provides an algorithmic reduction of nonlocal overhead and constitutes a promising building block for scalable DQC.

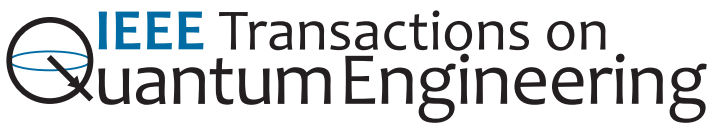

Finally, while our results focus on star networks, the same principles extend naturally to hierarchical multitier networks. In this setting, routers could cooperatively generate and manage entangled states across tiers, alleviating bottlenecks and improving scalability. A detailed analysis of coordination overheads and heterogeneous latencies is left for future work, but we stress that our method remains compatible with such topologies and should yield similar reductions in nonlocal depth.

## ACKNOWLEDGMENT

The authors thank Dr. Constantino Rodríguez for their many interesting and fruitful discussions. During the preparation of this work, the authors have used AI tools to revise grammar and readability. The authors take full responsibility for the content of the publication.